\documentclass[12pt,preprint]{aastex}

\usepackage{psfig}
\include{epsf}
\newcommand{\be}{\begin{equation}}
\newcommand{\ee}{\end{equation}}
\newcommand{\ba}{\begin{eqnarray}}
\newcommand{\ea}{\end{eqnarray}}
\newcommand{\bsf}[1]{\mbox{\begin{bfseries}\textsf{{#1}}\end{bfseries}}}

\shorttitle{Semi--linear lens inversion}
\shortauthors{Warren \& Dye}

\begin{document}


\title{Semi--linear gravitational lens inversion}


\author{S. J. Warren and S. Dye } \affil{Astrophysics Group, Blackett
Laboratory, Imperial College London, Prince Consort Road, London, SW7 2BW, UK}

\begin{abstract}

We describe a new method for analyzing gravitational lens images, for
the case where the source light distribution is pixelized. The method
is suitable for high resolution, high $S/N$ data of a multiply--imaged
extended source.  For a given mass distribution, we show that the step
of inverting the image to obtain the deconvolved pixelized source
light distribution, and the uncertainties, is a linear one. This means
that the only parameters of the non--linear problem are those required
to model the mass distribution. This greatly simplifies the search for
a min.$-\chi^2$ fit to the data and speeds up the inversion. The
method is extended in a straightforward way to include linear
regularization. We apply the method to simulated Einstein ring images
and demonstrate the effectiveness of the inversion for both the
unregularized and regularized cases.

\end{abstract}


\keywords{gravitational lensing}

\section{Introduction}
\label{sec:intro}

This paper is concerned with the problem of inversion of a
gravitationally--lensed image of an extended source, i.e. a galaxy
rather than a star or quasar. This problem is interesting because
lensed images of extended sources provide more information than images
of point sources, and so potentially are more useful for determining
the mass profiles in galaxies and clusters of galaxies. Also, because
of the magnification, one can measure structure in the light profile
of the source at enhanced resolution. In this paper we show how this
problem can be separated in a natural way into linear and non--linear
dimensions, and how this provides a number of advantages.

In this introduction we review solutions to the inversion problem and
introduce some of the terminology used in the remainder of the paper.
In all the solutions described here the mass in the lens is
parameterized. Nevertheless the analysis applies equally to a
pixelized mass distribution.

When presented with the lensed image of an extended source, the
unknowns to solve for are the source light profile and the lens mass
profile, and the uncertainties in these quantities. One approach to
this problem, suggested by Kayser and Schramm (1988), uses the fact
that regions of the source that are multiply--imaged have the same
surface--brightness. For a trial mass distribution, the method traces
image pixels to the source plane where the counts in different image
pixels mapping to the same source pixel are compared. The solution for
the mass is obtained by minimizing the dispersion in the image pixel
counts for such multiply--imaged source pixels. Kochanek et al. (1989)
successfully applied this approach to the inversion of the radio
Einstein ring MG1131+0456. The algorithm was refined by Wallington,
Kochanek, and Koo (1995) who applied it to the triply--imaged giant
arc in the galaxy cluster Cl 0024+1654.

The main shortcoming of this approach is that it does not deal with
the image point--spread--function (psf). If psf smearing of the image
(either instrumental or atmospheric) is significant, the light profile
of the source is not correctly recovered by backward tracing the
image, even if the mass distribution is exactly known. To deal with
the psf, a forward approach is needed i.e. one chooses a model for the
source light profile (parameterized or pixelized), and a model for the
mass (parameterized or pixelized), forms the image, convolves it with
the psf, and compares it to the actual image, adjusting the source and
lens models to minimize a merit function e.g. $\chi^2$.

An argument for choosing to parameterize rather than pixelize the
source light profile is that it forces the solution to be smooth.
Nevertheless, the source light profile may be complex. This is true,
for example, in the cases of MG1131+0456 and Cl\,0024+1654, cited
above. A large number of parameters might be required to provide a
satisfactory description. Without clues to the character of the source
it is extremely difficult to select the best parameterization i.e. the
one which provides a satisfactory fit with the smallest number of
parameters. In the most extreme example Tyson, Kochanski, \&
dell'Antonio (1998) used 232 parameters to model the source light
distribution of the galaxy lensed by the cluster Cl0024+1654.

If the source light profile is complex it is natural to consider
pixelizing the source, i.e. the counts in each pixel is a free
parameter. This removes the difficulty in finding a good
parameterization for the source, and thereby avoids any bias in the
fitted mass profile resulting from a poor choice. On the other hand,
due to the deconvolution, and because the pixels are independent, the
solution can be noisy. It is possible to achieve a smooth pixelized
solution by adding a suitable `regularizing' term to the merit
function. This term, if minimized on its own, would produce a smooth
source light profile. By adding this term to $\chi^2$ the final
solution involves a balance between obtaining the best fit to the
image (minimizing $\chi^2$), and obtaining a smooth source solution
(minimizing the regularizing term). Wallington, Kochanek, and Narayan
(1996) apply this approach to the case of the radio Einstein ring MG
1654+134. They use a maximum entropy approach i.e. the regularizing
term to be minimized is the negative of the entropy, the negentropy.
Labeling the counts in source pixel $i$ by $s_i$, the source plane
negentropy is $\sum_i s_i\ln(s_i)$, and the
merit function
they minimize is 
\be
\label{eq:eqa}
  G=\chi^2_{im}+\lambda\sum_i s_i\ln(s_i) 
\ee 
(here we have followed the notation in Press et al., 2001, \S 18.7).
The purpose of the multiplier $\lambda$ is to give more or less weight
to the negentropy term.

The inversion proceeds as follows: For a fixed value of $\lambda$, the
solution is determined by searching through the multi--parameter space
for the minimum of the merit function. The number of dimensions of the
parameter space to search is the sum of the number of source pixels
and the number of parameters used for the mass. This search is most
efficiently achieved with a pair of nested cycles. The inner (source)
cycle searches for the best source light profile for a fixed mass
profile. The outer (mass) cycle adjusts the mass profile. Outside this
cycle is a third ($\lambda$) cycle where the multiplier is
adjusted. Because the negentropy term acts to smooth the source, as
$\lambda$ increases, the $\chi^2_{im}$ term also increases, i.e. the
fit becomes worse. The principle for reaching the final solution
(e.g. Press et al., 2001, \S 18.4) is to start with $\lambda$ large,
then progressively reduce the weight of the regularizing term until
the $\chi^2_{im}$ becomes satisfactory. In other words the solution
has the smoothest source that provides a satisfactory fit to the
image. `Satisfactory' is usually interpreted as reaching the criterion
for the $\chi^2$ for the image
$\chi^2_{im}=min(\chi^2_{im})+\sigma(\chi^2_{im})$.  With three nested
cycles, the $\lambda$, mass, and source cycles, the routine can be
slow.

In this paper we describe a new technique which we suggest simplifies
and clarifies the problem in a number of ways. In purely formal terms,
the method is very similar to the maximum entropy method of Wallington
et al.: Algebraically, we simply replace the negentropy term in the merit
function (\ref{eq:eqa}) with a linear regularization term. However,
the insight we bring is to show that for a fixed mass distribution,
the minimization of the merit function is now a linear problem
i.e. can be solved by matrix inversion. In other words the source
cycle \---\ the innermost of the three minimization cycles \---\ is
eliminated. This has major benefits. In the first place the inversion
is much quicker, thereby allowing a more thorough search for the best
fit mass model. At the same time, the uncertainty of identifying the
true minimum has been removed. The method also greatly simplifies
calculation of the uncertainties, as we show below. More generally,
the formalism clarifies the essence of the problem: The source
parameters are linear dimensions and the mass parameters are
non--linear dimensions. For this reason we call the method
`semi--linear'.

At this point it is worth noting that, because of magnification and
multiple imaging, the number of constraints to the solution can be
much greater than the number of parameters to solve for. In this
respect the lens inversion problem differs from many inversion
problems encountered in astronomy (for example image
deconvolution). We find, as a consequence, that in many circumstances
the regularization term can be removed altogether. So the $\lambda$
cycle is also eliminated. The merit function is then just
$\chi^2_{im}$, and this is our starting point for the presentation of the
theory. For a fixed mass profile, the pixelized source light
distribution that produces the min.$-\chi^2_{im}$ fit is obtained by
linear inversion. The mass profile is then adjusted to find the
minimum of these min.$-\chi^2_{im}$ fits. The advantage, besides speed
(only the mass cycle remains), is that the solution is unbiased, since
there is no smoothing of the source.

The outline of the remainder of the paper is as follows. In
\S\ref{sec:theory} we explain the basic theory, demonstrating that for
a fixed mass profile, the problem of obtaining the source light profile
by $\chi^2$ minimization in the image plane is a linear one, and
obtaining the covariance matrix for the counts in the source
pixels. We then extend the basic theory to include a linear
regularization term. In \S\ref{sec:sims} we apply the method to a
realistic problem, assessing the performance for different psf widths,
and different source pixel sizes, with and without linear
regularization. In \S\ref{sec:conc} we provide a summary of the main
points, together with recommendations for applying the method.

\section{Theory}
\label{sec:theory}

In this section we present the theory of semi--linear inversion,
firstly without regularization (\S\ref{sec:theory.unreg}), and then
with regularization (\S\ref{sec:theory.reg}). In each sub--section we
begin with the case where the mass is fixed, and then treat the
general case, minimizing also on the mass parameters.

\subsection{Semi--linear inversion without regularization}
\label{sec:theory.unreg}

\subsubsection{Fixed mass: Eliminating the source cycle}
\label{sec:theory.unreg.fix}

Without any regularizing term, the merit function is simply
$G=\chi^2_{im}$. The basic problem is to find the counts in the source
pixels that, for a given mass distribution, minimize the merit
function $G$, i.e. give the best fit to the observed image. Pixels in
the source plane are labeled $i=1,I$. There is no restriction on how
the source plane is tessellated. In principle, the pixels could change
in both size and shape across the source region, which itself could be
of any shape. Pixels in the image plane are labeled $j=1,J$. It is
assumed in the following that the image pixels include counts from the
image of the lensed source only i.e. the images of any lensing
galaxies, and the mean sky count, have been subtracted. Also we suppose
that the data in each image pixel are independent i.e. are
characterized by the counts $d_j$, and dispersion $\sigma_j$, with no
covariance between pixels (appropriate for CCD data).

The inversion proceeds as follows: Choose the mass model parameters,
then, for each source pixel $i$, in turn, form the image for unit
counts (surface brightness) by appropriate ray tracing and convolution
with the known point spread function i.e. compute the counts in the
$i$th image $f_{ij},j=1,J$. The problem now is to combine these $I$
images with scalings $s_i,i=1,I$, to minimize $G$. These scalings are
the deconvolved intrinsic source surface--brightness distribution.

The problem is of a standard type. The merit function is written
\be
\label{eq:eqb}
  G=\chi^2_{im}=\sum_{j=1}^{J}
  \left[\frac{\sum_{i=1}^{I}s_if_{ij}-d_j}{\sigma_j}\right]^2.
\ee

Minimizing with respect to each of the source terms we have a set of
$I$ simultaneous equations of the form
\be
\label{eq:eqc}
  \frac{1}{2}\frac{\partial{G}}{\partial{s_i}}=0=\sum_{j=1}^{J}
  \left[\frac{f_{ij}\sum_{k=1}^{I}s_kf_{kj}-f_{ij}d_j}{\sigma_j^2}\right]
\ee
where the reason for the factor $\frac{1}{2}$ will soon become clear.
These equations may be written in matrix form
\be
\label{eq:eqd}
  \bsf{F}\bsf{S}=\bsf{D}.
\ee
Here \bsf{S} is a column matrix of length $I$ containing the elements
$s_i$, to be solved for. \bsf{F} is a symmetric $I\times I$ matrix,
with elements
$\bsf{F}_{ik}=\sum_{j=1}^{J}f_{ij}f_{kj}/\sigma_j^2$. Finally \bsf{D}
is a column matrix of length $I$ containing the elements
$\bsf{D}_i=\sum_{j=1}^{J}f_{ij}d_{j}/\sigma_j^2$.

The solution for the counts in the source pixels is then simply
obtained by matrix inversion
\be
\label{eq:eqe}
  \bsf{S}=\bsf{F}^{-1}\bsf{D}
\ee
thus eliminating the source cycle.

The solution for the errors has a particularly simple form. We seek the
covariance matrix for the source pixels. Noting that
\be
  \label{eq:eqf}
  \bsf{F}_{ik}=\frac{1}{2}\frac{\partial{^2G}}{\partial{s_i}\partial{s_k}},
\ee
we see that the matrix \bsf{F} is one half times the Hessian matrix of
$\chi^2_{im}$, which is to say that \bsf{F} is the {\em curvature
matrix} of the problem (Press et al., 2001, \S\S15.4, 15.5) \---\ this
was the reason for using the factor $\frac{1}{2}$ in equation
(\ref{eq:eqc}). We now show that the matrix $\bsf{C}=\bsf{F}^{-1}$ is
the required covariance matrix of $\bsf{S}$.

For independent image pixels, the covariance between source pixels $i$ 
and $k$ is given by
\be
  \label{eq:eqg}
  \sigma_{ik}^2=\sum_{j=1}^{J}\sigma_{j}^2\frac{\partial{s_i}}
  {\partial{d_j}}\frac{\partial{s_k}}{\partial{d_j}}.
\ee
Using equation (\ref{eq:eqe}) this becomes
\be
  \label{eq:eqi}
  \sigma_{ik}^2=\sum_{j=1}^{J}\sigma_{j}^2
  \sum_{l=1}^{I}\bsf{C}_{il}\frac{f_{lj}}{\sigma_j^2}
  \sum_{m=1}^{I}\bsf{C}_{km}\frac{f_{mj}}{\sigma_j^2}.
\ee
Multiplying this out gives
\be
  \label{eq:eqj}
  \sigma_{ik}^2=\bsf{C}_{ik}
\ee
as required.

We see that for the case of fixed mass, the covariance matrix for the
source pixel counts is computed in the inversion step, without the
need for further calculation. We shall refer to this $I\times I$
matrix as the {\em source covariance matrix} hereafter. Even though it
is not the complete solution for the source pixel errors (because the
mass parameters have been fixed), the source covariance matrix is
extremely useful, for example, in exploring different mass models and
pixelizations (\S\ref{sec:sims}).

It is worth noting that the semi--linear inversion solution, either
with or without regularization, differs in character from the maximum
entropy solution. With the semi--linear method the counts in any
source pixel are unbounded, so the best--fit value could be negative,
since some image pixels contain negative counts (i.e. are below mean
sky). With the maximum entropy method all source counts must be
positive. The semi--linear method provides the best estimate of the
counts in a source pixel, and the solution is satisfactory provided
the result is consistent with being positive. If the counts in any
source pixel are significantly negative (e.g. $<-4\sigma$) this would
indicate a bad mass model. This possibility of testing the suitability
of the mass model with a source--plane statistic can be viewed as an
extra positive feature of the semi--linear method.

\subsubsection{Mass cycle}
\label{sec:theory.unreg.mass}

The full solution proceeds by searching through the mass--distribution
parameter space, at each point minimizing $\chi^2_{im}$ by linear
inversion, to find the smallest of these min.$-\chi^2_{im}$ values,
the global minimum. Because the number of dimensions of the parameter
space for the non--linear search has been greatly reduced, it is now a
much simpler problem to locate the true minimum securely. 

The solution for the errors is more complicated than above, since we
have added in the non--linear mass dimensions. If there are $L$
parameters that describe the mass, labeled $m_l$, we need to form the
$(I+L)\times(I+L)$ (symmetric) curvature matrix of the problem. But
note that the majority of the terms, the $I\times I$ terms
$\frac{1}{2}\frac{\partial{^2G}}{\partial{s_i}\partial{s_k}}$, have
already been computed and are the elements of the matrix \bsf{F} at
the global minimum. The remaining terms, the $L$ rows (and columns) of
terms such as
$\frac{1}{2}\frac{\partial{^2G}}{\partial{m_l}\partial{m_n}}$, and
$\frac{1}{2}\frac{\partial{^2G}}{\partial{m_l}\partial{s_i}}$, can be
filled in by simple measurement of the shape of the $\chi^2_{im}$
surface about the global minimum. The covariance matrix for the mass
and source parameters is the inverse of this curvature matrix.  We
shall refer to this $(I+L)\times(I+L)$ matrix as the {\em full
covariance matrix} hereafter.

\subsection{Semi--linear inversion with regularization}
\label{sec:theory.reg}

\subsubsection{Fixed mass: Eliminating the source cycle}
\label{sec:theory.reg.fix}

The possibility of replacing the negentropy term in equation
(\ref{eq:eqa}) by a term (a linear regularization term) which
preserves the linearity of the min.$-\chi^2$ approach is made evident
by the linearity of equation (\ref{eq:eqc}) with respect to the source
parameters. Clearly we can form a merit function by adding to
$\chi^2_{im}$ any term $G_L$ which is a linear combination of terms
$s_is_k$
\be
  \label{eq:eqk}
  G_L=\sum_{i,k}a_{ik}s_is_k
\ee
since the partial differentials of these additional terms will also be
linear. One example of a linear regularization term is
$G_L=\sum_{i=1}^{I}s_i^2$. The choice of $G_L$ is discussed below.

Writing the merit function generally as
\be
  \label{eq:eql}
  G=\chi^2_{im}+\lambda G_L
\ee
then, following through the same analysis as in
\S\ref{sec:theory.unreg.fix}, the solution for the counts in the
source pixels can be written
\be
\label{eq:eqm}
  \bsf{S}=[\bsf{F}+\lambda\bsf{H}]^{-1}\bsf{D}.
\ee
We call the matrix \bsf{H} the regularization matrix. The elements of
\bsf{H} are 
\be 
  \label{eq:eqn}
  \bsf{H}_{ik}=\frac{1}{2}\frac{\partial{^2G_L}}{\partial{s_i}\partial{s_k}}.
\ee
For example, if the regularization term is $G_L=\sum_{i=1}^{I}s_i^2$,
then we have $\bsf{H}=\bsf{I}$, the identity matrix.

The form of $G_L$ should be chosen to penalize noisy solutions. The
choice $G_L=\sum_{i=1}^{I}s_i^2$, termed ``zeroth--order''
regularization in the literature, is one attempt to achieve
this. Other widely--used linear regularization terms include {\em
gradient} and {\em curvature} forms. These three regularization forms
correspond, loosely speaking, to the prejudice that the source light
profile is, respectively, approximately zero, constant, or planar (see
Press et al., 2001, \S18.5, for a detailed account of the theory of
linear regularization and its implementation). In practice, if
$\lambda$ is not too large, all three terms serve to smooth the source
in a rather similar way, and there is little to distinguish between
the solutions.

The gradient and curvature forms consider the differences between
counts in neighboring pixels. Until now we have used a
one--dimensional numbering scheme for the source pixels. In this case,
since we need to take account of the relative spatial locations of
pixels in the source plane we use coordinates $x, y$. The simplest
gradient term is
\be 
  \label{eq:eqo}
  G_L=\sum_{x,y}[s_{x,y}-s_{x+1,y}]^2+\sum_{x,y}[s_{x,y}-s_{x,y+1}]^2.
\ee 
Another form uses $[s_{x,y}-0.5(s_{x+1,y}+s_{x,y+1})]^2$. In forming
the sum it is necessary to translate the indices $x,y$ to the index
$i$, and then equation (\ref{eq:eqn}) is used to form the matrix
\bsf{H}. Note that zeroth--order regularization is computationally by
far the simplest method, since it does not involve this step of
translation of indices.

The simplest curvature form is
\ba 
  \label{eq:eqp}
  G_L=\sum_{x,y}[s_{x,y}-0.5(s_{x-1,y}+s_{x+1,y})]^2 \nonumber \\
  +\sum_{x,y}[s_{x,y}-0.5(s_{x,y-1}+s_{x,y+1})]^2.
\ea
Another form uses
$[s_{x,y}-0.25(s_{x-1,y}+s_{x+1,y}+s_{x,y-1}+s_{x,y+1})]^2$.

The source covariance matrix for the regularized case is fortunately
only slightly more difficult to compute than for the unregularized
case. Writing $\bsf{R}=[\bsf{F}+\lambda\bsf{H}]^{-1}$, and following
the same line of reasoning as in \S\ref{sec:theory.unreg.fix}, we
obtain the analogous equation to equation (\ref{eq:eqi})
\be
  \label{eq:eqq}
  \sigma_{ik}^2=\sum_{j=1}^{J}\sigma_{j}^2
  \sum_{l=1}^{I}\bsf{R}_{il}\frac{f_{lj}}{\sigma_j^2}
  \sum_{m=1}^{I}\bsf{R}_{km}\frac{f_{mj}}{\sigma_j^2}.
\ee
Multiplying out we obtain
\be
  \label{eq:eqr}
  \sigma_{ik}^2=\bsf{R}_{ik}-
  \lambda\sum_{l=1}^I \bsf{R}_{il}[\bsf{RH}]_{kl}.
\ee

\subsubsection{Mass cycle}
\label{sec:theory.reg.mass}

The procedure for the full solution is the same as for the
unregularized case. One searches through the mass--distribution
parameter space, at each point minimizing $G$ by linear inversion, to
find the smallest of these min.$-G$ values, the global minimum. In the
regularized case there is no simple solution for the full covariance
matrix however. In the unregularized case, we were able to use the
fact that the inverse of the full curvature matrix is the full
covariance matrix. But in the regularized case this is no longer true
since we are minimizing $G=\chi^2_{im}+G_L$. Instead, an alternative
approach must be used, for example, a Monte Carlo method which inverts
an ensemble of realisations of the image by adding noise to the
original image.

\section{Simulations}
\label{sec:sims}

In this section we apply the semi--linear inversion method to a
realistic test problem. To validate the linear inversion step, we
begin with the case of fixed mass. We quantify the effectiveness of
the method under variations of the image $S/N$, psf width, and source
pixel size, for both the unregularized and regularized cases. We then
present an analysis of the full problem, allowing the mass parameters
to vary. Finally we debate the advantages of the unregularized and
regularized approaches, for different practical applications.

\subsection{Test problem}
\label{sec:sims.test}

To make the computations more useful we have based our investigation
on a realistic simulation of a deep image of an Einstein ring
gravitational lens system observed with the Advanced Camera for
Surveys (ACS) aboard HST. The camera has a pixel size of
$0.05\arcsec$.  We have used cosmological parameters $\Omega_m=0.3$,
$\Omega_{\Lambda}=0.7$. The lens is placed at $z=0.3$ and the source
lies at $z=3.0$.

Figure 1 illustrates the test problem. The lens (not shown) is
modelled as a singular isothermal ellipsoid with one--dimensional
velocity dispersion $260$ km s$^{-1}$ and ellipticity
$e=1-b/a=0.25$. The semi--major axis of the lens is aligned at
$40^{\circ}$ counterclockwise from the vertical. The Einstein angle is
$\theta_E=4\pi\sigma^2D_{ds}/(c^2D_s)=1.58\arcsec$. The source, shown
in the top left panel, is contained within a square of size
$0.75\arcsec$ and is modelled as two circular sources of Gaussian
profile, binned in $0.05\arcsec$ pixels, the same as the image pixel
size. The peak surface brightness of each source is 1.0, in arbitrary
units.  One source lies inside the inner caustic, while the second
source straddles the inner caustic. This source configuration
resembles that inferred for the gravitational lens $0047-2808$ (Wayth
et al., 2003). To create a realistic ACS simulation the image was
formed by ray tracing, then convolved with the point spread function,
and noise added (in the manner described in the following
paragraph). For simplicity we modelled the psf as a Gaussian, and
chose FWHM $0.08\arcsec$ which is the resolution of a
diffraction--limited telescope of the same diameter as HST, at a
wavelength $\lambda=800$nm. Because of the slight undersampling, the
convolution is made on a sub--pixelized grid and then binned up to the
full pixel size.

The data pixels used for the inversion were the 3626 pixels within the
annulus in the image plane marked in the figure. This annulus is
defined by the region covered by imaging the entire source
plane.\footnote{The region of the central image should also be
included for non--singular mass models.} An important point to note is
that the analysis region must at least cover this annulus, otherwise
the counts in some source pixels will be unconstrained and the
inversion will fail. A larger region may be used, but if it becomes
too large the usefulness of the $\chi^2$ statistic is diminished, as
then most of the pixels are in the background. The final step in the
simulation is to apply uniform Gaussian random noise over the image
plane. The noise level is defined in terms of the total $S/N_{im}$
integrated over the annulus. The same noise realisation was used for
all the simulations, but scaled in order to vary $S/N_{im}$.

The upper middle panel shows the final simulated ACS image. This
image, with source pixel size $0.05\arcsec$, $S/N_{im}=60$, and psf
FWHM$=0.08\arcsec$, is the reference test problem to invert. We later
vary these three parameters. The parameters of the different models we
have run are listed in Table \ref{table1}. Col. (1) provides the
simulation number. The reference problem is numbered 1. Col. (2) gives
the source pixel size in arcsec, col. (3) the summed $S/N$ in the
image, and col. (4) the psf FWHM in arcsec. Col. (5) is a label U or R
depending on whether the inversion was unregularized or
regularized. Then, in the case of regularized inversion, col. (6)
provides the degree of regularization, quantified by the parameter
$N_\lambda$, which is the increase of $\chi^2_{im}(\nu)$ from the
minimum in units of the standard deviation $\sigma(\chi^2_{im}(\nu))$.
Recalling the discussion in \S1, a value $N_\lambda=1$ in this column
corresponds to the usual criterion for the maximum allowed degree of
regularization. The other columns are explained in the next section.

\clearpage
  
\begin{figure}
\label{fig:lowres.unreg}
\plotone{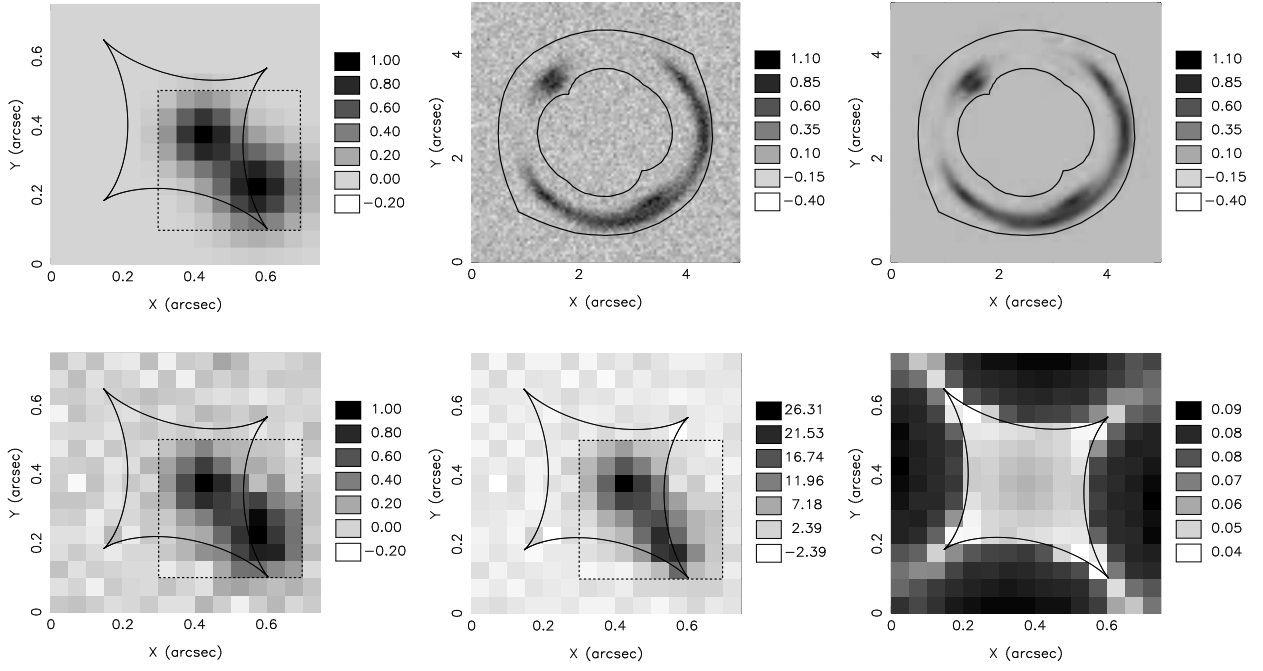}
\caption{This plot shows the unregularized solution for the reference
problem, line (1), Table 1. The source plane, top left panel and
bottom row, is $0.75\arcsec\times0.75\arcsec$ with $0.05\arcsec$
pixels, and is centered on the optic axis. The source comprises two
circular Gaussian components and is shown top left. Also marked is the
line of the inner caustic for the isothermal ellipsoid lens. The
image, convolved with the psf, FWHM $0.08\arcsec$, and with noise
added, is shown upper middle. The image pixel size is $0.05\arcsec$
and the image box size is $5.0\arcsec\times5.0\arcsec$.  The lower
left panel is the source light distribution reconstructed from the
image by semi--linear inversion without regularization. The upper
right panel is the image of this source, convolved with the psf. The
lower right panel displays the $1\sigma$ uncertainty for the source
pixels, and the lower middle panel is the source $S/N$ image. The
dotted square is the region over which $S/N_{so}$ is measured. In this
and the following two figures counts in pixels in both the image and
source plane are in units of surface brightness.}
\end{figure}

\clearpage

\begin{table}
\footnotesize
\centering
\begin{tabular}{|rcccccccrccr|}
\hline
%
%
(1) & (2) & (3) & (4) & (5) & (6) & (7) & (8) & (9) & (10) & (11) &
(12) \\
sim. &
source                           & 
$S/N_{im}$     & 
psf	           &  
U/R &
$N_\lambda$ &
$\chi^2_{im}(\nu)$                    &  
$\chi^2_{so}(\nu)$                     &   
$S/N_{so}$      & 
$|\Delta s/\sigma|$          & 
$\Delta s_{rms}$             &
note   \\
no. & pix. size & & FWHM & & & & & & max. & & \\
 & $''$ & & $''$ & & & & & & & & \\
\hline
%
 1 & 0.050 & 60.0 & 0.08 & U &   & $0.956\pm0.024$ & $1.088\pm0.094$ &
 79.9 & 2.80 & 0.082 & Fig. 1 \\
 2 & 0.050 & 30.0 & 0.08 & U &   & $0.956\pm0.024$ & $1.088\pm0.094$ & 
40.6 & 2.80 & 0.163 & \\
 3 & 0.050 & 60.0 & 0.00 & U &   & $0.958\pm0.024$ & $1.052\pm0.094$ & 
111.2& 2.64 & 0.040 & \\
 4 & 0.050 & 60.0 & 0.16 & U &   & $0.956\pm0.024$ & $1.090\pm0.094$ & 
24.6 & 2.73 & 0.252 & \\
 5 & 0.025 & 60.0 & 0.08 & U &   & $0.952\pm0.027$ & $1.003\pm0.047$ &
 20.2 & 2.84 & 0.634 & Fig. 2 \\ \hline
 6 & 0.050 & 60.0 & 0.08 & R & 1 & $0.980\pm0.024$ & $1.138\pm0.094$ & 
111.8& 3.02 & 0.031 & \\
 7 & 0.050 & 60.0 & 0.08 & R & 2 & $1.004\pm0.024$ & $1.461\pm0.094$ & 
120.8& 4.35 & 0.028 & \\
%
8 & 0.025 & 60.0 & 0.08 & R & 1 & $0.979\pm0.027$ & $1.003\pm0.047$ &
64.5 & 3.14 & 0.242 & Fig. 3 \\
9 & 0.025 & 60.0 & 0.08 & R & 3 & $1.033\pm0.027$ & $1.003\pm0.047$ & 
86.6& 3.40 & 0.123 & Fig. 3 \\
10 & 0.025 & 60.0 & 0.08 & R & 5 & $1.088\pm0.027$ & $1.004\pm0.047$ & 
98.2& 3.34 & 0.070 & \\
\hline
\end{tabular}
\caption{\normalsize Dependence of reconstruction performance on
source plane pixel size, simulated ring image noise, psf width, and
degree of regularization.}
\label{table1}
\end{table}

\clearpage

\subsection{Fixed mass}
\label{sec:sims.fix}

The main purpose of the simulations is to illustrate the linear
inversion step, the `inner cycle', i.e. that part of the semi--linear
inversion method that differs from previous methods. For this reason
in this sub--section we fix the mass parameters at the input
values. The image inversion, therefore, is achieved in a single step
using equations (\ref{eq:eqe}) and (\ref{eq:eqm}), for the
unregularized and regularized cases respectively, and using equations
(\ref{eq:eqj}) and (\ref{eq:eqr}) for the source covariance matrix. We
consider the full problem, solving also for the mass parameters in
\S\ref{sec:sims.mass}.

\subsubsection{Unregularized inversion}
\label{sec:sims.fix.unreg}

The unregularized inversion of the reference problem is provided in
the remaining panels of Figure 1. The lower left panel shows the
reconstructed source and the upper right panel shows the image of the
reconstructed source convolved with the psf, i.e. the
min.$-\chi^2_{im}$ model fit to the simulated image. The bottom right
panel shows the source $\sigma$ image i.e. the standard deviation in
each pixel. This provides a visual impression of the uncertainties
\---\ note how the region of lowest $\sigma$ is bounded by the inner
caustic. However the whole covariance matrix is required for a proper
interpretation of the results.  The lower middle panel is the source
$S/N$ image. In all the Figures $1-3$, for the source $\sigma$ and
$S/N$ images the grayscale covers the full range of numbers in the
panel. For the other panels the same grayscale range is used in each
figure, to allow comparison of the relative noise levels.
 
We measure several quantities to assess the quality of the inversion,
listed in the remaining columns of Table \ref{table1}. The reduced
$\chi^2$ in the image plane, $\chi^2_{im}(\nu)$ is provided in
col. (7). The quoted uncertainty is given by $\sqrt{2/\nu}$ where
$\nu$ is the number of degrees of freedom i.e. the number of image
pixels (3626) minus the number of source pixels (225 or 900).  The
reduced $\chi^2$ in the source plane, $\chi^2_{so}(\nu)$, and its
uncertainty, is provided in col. (8). To account for the covariance
terms this is computed using
\be
\label{eq:eqs}
\chi^{2}_{so}=\sum_{i,k} \Delta s_i \bsf{C}^{-1}_{ik} \Delta s_k
= \sum_{i,k} \Delta s_i \bsf{F}_{ik} \Delta s_k 
\ee 
where $\Delta s_i$ is the residual in the $i$th pixel.  Here the
number of degrees of freedom is the number of source pixels. Col. (9)
provides the $S/N$ summed over the small box in the source plane shown
in Figure 1. The noise is computed as the square root of the sum
of the elements in the covariance matrix, formed by stripping out from
the source covariance matrix $\bsf{C}$ the rows and columns
corresponding to the pixels in the box. Col. (10) provides the
absolute value of the significance $\Delta s/\sigma$ of the worst--fit
source pixel, and col. (11) lists the $r.m.s.$ of the residuals in the
source plane.

The results for the reference problem, line (1) in Table \ref{table1},
are all satisfactory: The reduced $\chi^2$ values in the image and
source planes are both consistent with $1.0$, and the significance of
the worst pixel $2.80\sigma$ (col. 10) is not unexpected given that
there are 225 source pixels. The summed $S/N$ in the source box is an
improvement on $S/N_{im}$. This might be expected since the box is
restricted to the small region of the source plane containing nearly
all the signal. At the same time it shows that the $S/N$ is not
greatly degraded by amplification of noise in the deconvolution
step. We return to this issue below. We interpret these results as
meaning that the inversion has succeeded and produced the correct
solution to the well--posed problem of finding the source--pixel
counts that give the best fit to the image.

In simulation 2 we doubled the noise in the image plane. Comparing
lines (1) and (2) in the table, the effect of this is to double the
noise in the source plane (col. 11), and so halve the S/N of the
detected source (col. 9), as expected.

\clearpage

\begin{figure}
\label{fig:hires.unreg}
\plotone{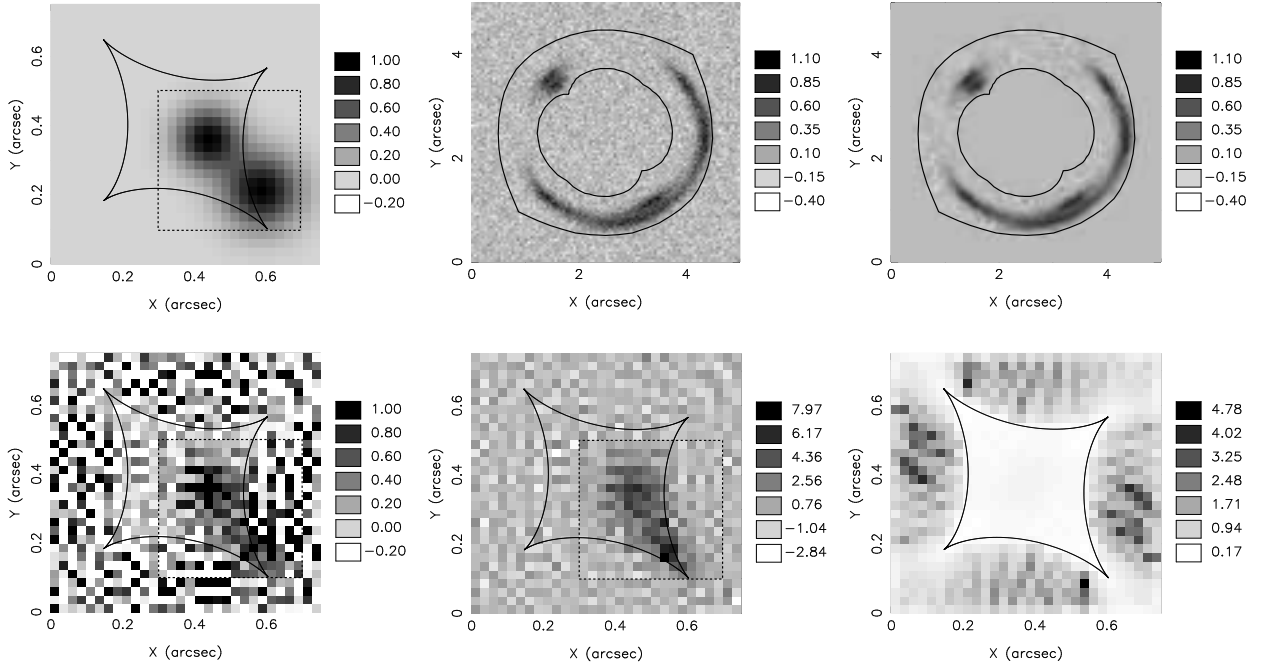}
\caption{The plot shows the unregularized solution for the same
problem as in Figure 1, but with source pixels half as large, and
corresponds to line (5), Table 1.  The source plane, top left panel
and bottom row, is $0.75\arcsec\times0.75\arcsec$ with $0.025\arcsec$
pixels, and is centered on the optic axis. The source comprises two
circular Gaussian components and is shown top left. Also marked is the
line of the inner caustic for the isothermal ellipsoid lens. The
image, convolved with the psf, FWHM $0.08\arcsec$, and with noise
added, is shown upper middle. The image pixel size is $0.05\arcsec$
and the image box size is $5.0\arcsec\times5.0\arcsec$.  The lower
left panel is the source light distribution reconstructed from the
image by semi--linear inversion without regularization. The grayscale
range is the same as in Figure 1. The reconstruction is poor, because
the source pixel size is too small. The upper right panel is the image
of this source, convolved with the psf. The lower right panel displays
the $1\sigma$ uncertainty for the source pixels, and the lower middle
panel is the source $S/N$ image. The dotted square is the region over
which $S/N_{so}$ is measured. In each of Figures 1--3, counts in
pixels in both the image and source plane are in units of surface
brightness.}
\end{figure}

\clearpage

%
%

{\em Variation of psf FWHM.} We have investigated the effect of
varying the width of the psf. Line (3) provides the results for no
psf, and line (4) provides the results for a psf FWHM of
$0.16\arcsec$, double the reference value. Comparison of lines (1),
(3), and (4) shows that as the psf FWHM increases the noise in the
source plane, col. (11), increases and the source detection $S/N$,
col. (9), decreases.  This is as expected: In Fourier space the effect
of the psf is to suppress the amplitude of the power spectrum of the
source for large wave numbers. Therefore in the deconvolution process
the noise on these scales is amplified. As the psf is broadened the
power suppression is greater, and so the noise amplification in the
deconvolution step is greater. The reduction in $S/N_{so}$ from 111.2
(line 3) to 79.9 (line 1), in going from no psf to psf FWHM of
$0.08\arcsec$, is quite modest. This demonstrates that satisfactory
inversion of ACS images using $0.05\arcsec$ source pixels is possible
without regularization.

Regardless of the degree of amplification of noise the various
statistical quantities in cols (7)--(11) of Table 1, lines (1)--(4),
are all reasonable. This shows that in these cases the inversion is
well behaved, and in none of the cases is the matrix ${\bsf F}$
singular. This contrasts with the usual inversion problem, for example
image deconvolution. With image deconvolution the number of parameters
to solve for (the counts in the deconvolved image pixels) is typically
the same as or greater than the number of constraints (the number of
image pixels). In lensing, because of magnification, the number of
image pixels may be much greater than the number of source
pixels. This suggests, further, that in regions where the
magnification is greatest it would be possible to use source pixels
smaller than the image pixels. We consider this issue below.

The results of these first four simulations indicate that, in some
circumstances, provided the psf is not too broad, unregularized
semi--linear inversion provides a useful solution.

{\em Variation of source pixel size.} Figure 2 shows the same problem
as Figure 1 but with $0.025\arcsec$ source pixels rather than
$0.05\arcsec$ pixels. The results are summarized in line (5) of Table
\ref{table1}. The quality of the reconstruction, lower left panel, is
now dramatically worse, and outside the central region is clearly
unsatisfactory. (The grayscale range of this panel is the same as in
Figure 1.) Compared to line (1) the noise in the source (col. 11) has
risen by a factor 8, whereas intuitively one would expect only a
factor 2 increase (4 times as many pixels). This is indicative of
large amplification of noise, because the psf has suppressed the
signal on these scales. This is a consequence of the fact that the
separation in the image plane of the images of two adjacent source
pixels is smaller than the psf size. Put another way, a resolution
element in the image plane, traced back to the source plane, is
oversampled by the source pixel size. The source covariance matrix now
contains large, predominantly negative, covariance terms which
correspond to the odd/even appearance in the outer regions of the
source plane.

The high noise level in the outer parts of the source--plane belies
the usefulness of this image. In fact the source is strongly detected,
albeit at reduced $S/N_{so}=20$, even though not readily apparent to
the eye. The source is clearly visible in the $S/N$ image, however. At
the same time the $\chi^2$ values in both the image and source planes
remain satisfactory. Because of the larger magnification, the
reconstruction is much better within the caustic line. This suggests
it would be advantageous to use a variable pixel size across the
source plane. For example, with reference to Figure 2, a scheme where
the pixel size is $0.05\arcsec$ outside the caustic and $0.025\arcsec$
inside might be appropriate. We need to identify a criterion for
choosing the pixel size that avoids the excessive amplification of
noise evident in Figure 2. There are clearly three variables which
determine the minimum source pixel size: The image pixel size, $a$,
the psf FWHM, $b$, and the magnification, $c$. We have had some
success with a scheme which relates the source pixel size to the
variable $max(a,b/2)/c^{1/2}$. The results will be reported elsewhere
(Dye and Warren, in prep.).

\clearpage

\begin{figure}
\label{fig:hires.reg1}
\plotone{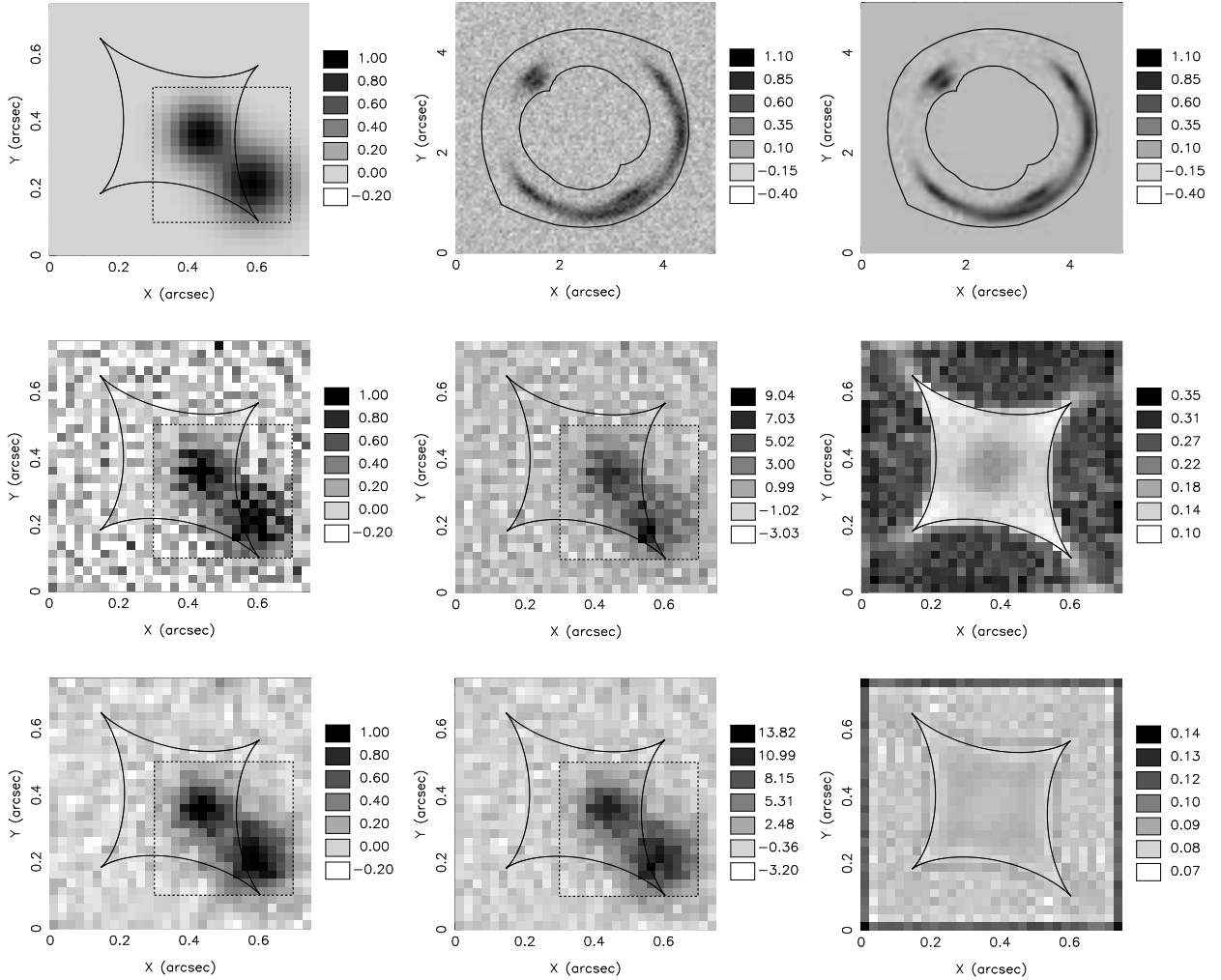}
\caption{\footnotesize The plot shows regularized solutions for the same problem as
in Figure 2, with different degrees of regularization. The middle row
is for $N_\lambda=1$ (corresponding to line (8) of Table 1), and the
bottom is for $N_\lambda=3$ (line (9) of Table 1). The source plane,
top left panel, and middle and bottom rows, is
$0.75\arcsec\times0.75\arcsec$ with $0.025\arcsec$ pixels, and is
centered on the optic axis. The source comprises two circular Gaussian
components and is shown top left. Also marked is the line of the inner
caustic for the isothermal ellipsoid lens. The image, convolved with
the psf, FWHM $0.08\arcsec$, and with noise added, is shown upper
middle.  The image pixel size is $0.05\arcsec$ and the image box size
is $5.0\arcsec\times5.0\arcsec$.  The left middle panel is the source
light distribution reconstructed from the image by semi--linear
inversion with regularization, $N_\lambda=1$. The solution is much
less noisy than the unregularized solution, Figure 2. The upper right
panel is the image of this source, convolved with the psf. The middle
right panel displays the $1\sigma$ uncertainty for the source
pixels. The center panel of the middle row
is the source $S/N$ image. The bottom row is the set of corresponding
source-plane images for the case $N_\lambda=3$. Note the larger errors
in the outermost band in the bottom right panel. This is a
consequence of the choice of a gradient regularization term, since
these pixels have fewer neighbours. The dotted square is
the region over which $S/N_{so}$ is measured. In each of Figures 1--3,
counts in pixels in both the image and source plane are in units of
surface brightness.}
\end{figure}

\clearpage

\subsubsection{Regularized inversion}
\label{sec:sims.fix.reg}

We now include linear regularization in the inversion. All the results
reported here used the gradient form, equation (\ref{eq:eqo}). The results
are quite similar for the different linear regularizing schemes
described in \S\ref{sec:theory.unreg.fix}, however.

Lines (6) and (7) in Table \ref{table1} are the results for the
reference problem with different degrees of regularization.  As the
regularizing term increases, the source becomes smoother and
$\chi^2_{im}$ increases.  Line (6) is for $N_\lambda=1$. Comparing
line (6) to line (1) we see that the effect of regularization is to
suppress the noise in the reconstructed source and to increase
substantially the source $S/N$, col. (9). This is at the expense of a
poorer match to the true source light profile, as measured by cols (8)
and (10). For $N_\lambda=2$ the agreement with the input source is no
longer acceptable.

In lines (8) to (10) we provide solutions for source pixel size
$0.025\arcsec$, and psf FWHM of $0.08\arcsec$, and different degrees
of regularization, $N_\lambda=1, 3, 5$. These results compare directly
to the unregularized solution to the same problem, line (5). The
solutions for simulation no. 8, $N_\lambda=1$, and no. 9,
$N_\lambda=3$, are shown in Figure 3. The visual improvement,
comparing the sequence of Figure 2 (unregularized), Figure 3 middle
row (regularzsed, $N_\lambda=1$), and Figure 3 bottom row (regularized,
$N_\lambda=3$), is dramatic.

Comparing lines (8) to (10) against line (5) we see that, again,
regularization successfully suppresses noise, increasing the $S/N$ of
the detection of the source.  As $N_\lambda$ increases, in this case
$\chi^2_{so}$ increases only very slowly, much more slowly than in the
case for larger pixels. This is partly due to the fact that we chose a
smooth source, and the results would be different for a source with
more small--scale structure. Nevertheless, it indicates that the
standard criterion for the degree of regularization to apply,
$N_\lambda=1$, is somewhat arbitrary.

To summarise this sub--section, using a realistic problem, we have
validated the theory of the linear inversion step set out in
\S\ref{sec:theory}. This is the step that differs from the
maximum--entropy method of Wallington et al. (1996), and therefore is
the main point of the paper.

\subsection{Mass cycle}
\label{sec:sims.mass}

In the present sub-section we report the results of solving the complete
problem i.e. determining both the mass profile and the source light
distribution.

We first consider the unregularized case. Referring back to the
example of Figure 1, the problem is to invert the image at upper
middle. The free parameters are the five parameters describing the mass:
$x$, $y$, ellipticity, position angle, and velocity dispersion. We
searched through the parameter space to find the min.$-\chi^2$ fit. At
the minimum the matrix ${\bsf F}$ supplies most of the terms of the
curvature matrix. Following the precepts of \S
\ref{sec:theory.unreg.mass}, the remaining terms were filled in by
measuring the relevant second partial derivatives of the $\chi^2$
surface. We found the surface to be completely smooth and parabolic
near the minimum. The full curvature matrix was inverted to obtain the full
covariance matrix for all the parameters, the mass terms as well as
the counts in the source pixels. We found the input mass parameters
were correctly recovered to within the uncertainties. We checked the
full covariance matrix against the results of Monte--Carlo simulations
and found excellent agreement. This confirms that, provided the chosen
source pixel size is not too small, the unregularized semi--linear
inversion method is a practical solution to the problem of inversion
of a gravitational lens image with a resolved source.

We also compared the terms in the {\em source covariance matrix}
${\bsf C}={\bsf F}^{-1}$ against the corresponding terms in the {\em
full covariance matrix}. The differences are relatively
small. Therefore the matrix ${\bsf C}$, at the global minimum,
provides an approximation to the true source--pixel errors that may be
very useful in the exploration stage, when considering different mass
models and different pixelizations.

In the regularized case we found, generally, that the procedure
converged more rapidly than in the unregularized case. Regularized
inversion can produce solutions which are not true representations of
the source (\S \ref{sec:sims.disc}, Table 1). Nevertheless, we found,
in contrast, that the solution for the mass parameters is very
insensitive to the degree of regularization. In the regularized case,
the curvature of the merit function cannot be used to obtain the full
covariance matrix so that an alternative approach such as a Monte Carlo
method must be adopted (\S\ref{sec:theory.reg.mass}). So the source
covariance matrix, equation (\ref{eq:eqr}), at the global minimum, is
particularly useful here as an approximation to the true source--pixel
errors.

\subsection{Regularized {\em vs} unregularized}
\label{sec:sims.disc}

We now include a debate on the relative advantages of the
unregularized and regularized approaches. This may seem surprising
given the excellent results achieved with regularization (comparing
Figures 2 and 3). The weakness of the unregularized inversion is that
in deconvolving the psf, the noise at large wavenumbers is
boosted. The regularization term in the merit function imposes
smoothness on the solution. In effect, the deconvolution (division in
Fourier space) is limited to the smaller wave numbers. However, this
means that any real structure in the source at large wavenumbers is
also suppressed. We are imposing a prejudice that the source is smooth
and this might not be justified (see comments in \S1). Regularization
introduces $+$ve covariance between adjacent pixels, forcing the
counts to be similar. The regularized solution, then, is not so
different to the unregularized solution with larger pixels. In this
respect, it is interesting to compare the unregularized solution with
$0.05\arcsec$ source pixels (Figure 1), with the $N_\lambda=1$
regularized solution with $0.025\arcsec$ source pixels (Figure
3). Noting that the two values of the source $S/N$ are quite similar
(79.9, 64.5, respectively, Table \ref{table1}), it is a debatable point
whether there is more information in the latter figure.

A further point to note is that the regularized inversion can produce
solutions which are satisfactory in terms of the fit to the image but
which are not true representations of the source in the sense that
the $\chi^2_{so}$ is unsatisfactory. For example in line (7), Table
\ref{table1}, both $\chi^2_{so}$ and $|\Delta s/\sigma|$ are
unsatisfactory. Measured by the same statistics, none of the
unregularized inversions in the Table is unsatisfactory. The
unregularized inversion gives a noisier but unbiased solution for the
source light distribution, while the regularized inversion gives a
smoother but biased solution.

Despite regularization biasing the source, in the full mass cycle we
find that the minimized mass parameters show little sensitivity to the
degree of regularization. Furthermore, the regularized solution has
the advantage that it converges more quickly. We discuss the practical
significance of these two points in \S\ref{sec:conc}. Associated with
this is the fact that regularization allows source pixel sizes of
almost any size, unlike the unregularized case when pixel size must be
chosen carefully. This can yield further speed advantages in the
initial stages of an analysis, before the solution is refined.

Overall we consider there are important advantages to using both
regularized and unregularized inversion in exploring the solution to a
particular problem, and the choice will depend on the question being posed
and any a priori knowledge concerning the source. Perhaps equally
importantly, however, it makes sense to match the source pixel size to
the data information content in terms of the $S/N$ at different
wavenumbers, or, in other words, to vary the pixel size depending on
the magnification, as suggested in \S\ref{sec:sims.fix.unreg}.

\section{Summary and recommendations}
\label{sec:conc}

We have developed a new method for the inversion of
gravitationally--lensed images of extended sources for the case where
the source light profile is pixelized. The method separates the linear
dimensions of the problem (the counts in the source pixels) from the
non--linear dimensions (the mass parameters). The method has been
extended in a natural way to allow linear regularization of the
inversion. The core of the routine is the procedure for inverting an
image given a fixed mass profile. We have shown that this step,
including deconvolution of the psf, with or without regularization, is
a linear one. Since this step is usually achieved by searching the
source parameter space for the merit--function minimum, the solution
is reached much more quickly. The non--linear part of the problem has
been reduced to the search for a minimum in the space of the mass
parameters only. In the case of unregularized inversion, the full
covariance matrix for all the (source$+$mass) parameters can be
obtained very quickly. In the case of regularized inversion, a useful
approximation to the covariance matrix for the source counts is
obtained very simply, but Monte Carlo methods are needed to obtain the
full covariance matrix.

How the semi--linear method should be applied in practice depends on
the problem posed. If one is interested in the quantitative details of
the source light profile, for example, whether some apparent feature
is real, then we recommend the unregularized solution. This is because
regularization produces source profiles which are too smooth.  Without
regularization, an optimal source pixel size should be chosen. Having
too large a source pixel may cause interesting detail to be
lost. However, if the source pixel size is too small, the inverted
image may have low $S/N$ because of amplification of noise in the
deconvolution step.  If, on the other hand, one is interested only in
the mass parameters, a regularized solution would be the appropriate
choice: The mass parameters are rather insensitive to the degree of
regularization and one benefits from an increase in inversion speed.

Another consideration is that pixelizing the source uses a large
number of parameters. As a rule, one is interested in finding the
model with the smallest number of parameters that provides a
satisfactory fit to an image. Therefore, in many cases, one might
simply use the semi--linear method of inversion to provide an image of
the source to guide the choice of parameterization. Here, again, the
regularized solution might be the preferred option.

In general, because it is so much easier to implement (\S
\ref{sec:theory.reg.fix}), we recommend using zeroth--order
regularization. Nevertheless, other considerations may override
simplicity. The zeroth--order regularization term, in common with the
maximum--entropy regularization term, is a local measure, independent
of the counts in adjacent pixels. This can be an advantage or a
disadvantage, depending on the actual light profile in the source.

In the simulations presented here, we have used square source pixels
which form a regular grid. However, since the resolution across the
image plane is fixed while the magnification varies, the resolution
across the source varies. Therefore, to maximize the information
content in the reconstruction of the source it is necessary to use a
variable source pixel size. We will present an analysis of
semi--linear inversion with variable source pixel size in a future
paper (Dye and Warren, in prep.).

\acknowledgments
We have benefited from discussions with Paul Hewett, Geraint Lewis,
Leon Lucy, and Randall Wayth.

\end{document}